\documentclass{emulateapj}
\usepackage{amsmath}
\usepackage{graphicx,natbib}
\usepackage{hyperref}
\def\Snospace~{\S{}}

\def\correct{\bgroup\let\next}
\newcounter{species} 
\def\ion#1#2{\setcounter{species}{#2}#1$\;${\scriptsize\Roman{species}}\relax}

\newcommand{\msun}{\;\rm{M_\sun}}
\newcommand{\aox}{\alpha_{OX}}
\shorttitle{EM Emission From Tidally Disrupted WD}
\shortauthors{Clausen \& Eracleous}
\begin{document}

\title{Probing Intermediate Mass Black Holes With Optical Emission Lines from Tidally Disrupted White Dwarfs}
\author{Drew Clausen and Michael Eracleous}
\affil{Department of Astronomy and Astrophysics and Center for Gravitational Wave Physics, 525 Davey Lab, The Pennsylvania State University, University Park, PA 16802, USA}

\begin{abstract}
We calculate the emission line spectrum produced by the debris released when a white dwarf (WD) is tidally disrupted by an intermediate-mass black hole (IMBH; $M\sim 10^{2}-10^{5}\msun$) and we explore the possibility of using the emission lines to identify such events and constrain the properties of the IMBH.  To this end, we adopt and adapt the techniques developed by Strubbe \& Quataert to study the optical emission lines produced when a main sequence (MS) star is tidally disrupted by a supermassive black hole.  WDs are tidally disrupted outside of the event horizon of a $< 10^{5}\msun$ black hole, which makes these tidal disruption events good signposts of IMBHs.  We focus on the optical and UV emission lines produced when the accretion flare photoionizes the stream of debris that remains unbound during the disruption.  We find that the spectrum is dominated by lines due to ions of C and O, the strongest of which are \ion{C}{4} $\lambda$1549 at early times and [\ion{O}{3}]  $\lambda$5007 at later times. Furthermore, we model the profile of the emission lines in the [\ion{O}{3}] $\lambda\lambda$4959, 5007 doublet and find that it is highly asymmetric with velocity widths of up to $\sim 2500 \rm{\;km\;s^{-1}}$, depending on the properties of the WD-IMBH system and the orientation of the observer. Finally, we compare the models with observations of X-ray flares and optical emission lines in the cores of globular clusters and propose how future observations can test if these features are due to a WD that has been tidally disrupted by an IMBH.  
\end{abstract}

\section{Introduction}
\label{intro}
A star is tidally disrupted when its orbit brings it too close to a black hole (BH).   If the two bodies are separated by less than the tidal disruption radius, $R_{T}\sim R_\star(M_{BH}/M_\star)^{1/3}$, \correct{where $R_{\star}$ is the radius of the star and $M_{BH}$ and $M_{\star}$ are the masses of the BH and star, respectively,} the star is ripped apart because the tidal acceleration across its diameter exceeds its self gravity.  \citet{Frank:1976} calculated the rate at which a BH in the core of a cluster or galaxy would disrupt and ingest \correct{main sequence (MS)} stars and suggested that the associated ultraviolet (UV)/X-ray flare would mark the presence of an otherwise undetectable BH.  This is possible because after the star is disrupted, about half of the debris becomes bound to the BH.  As this bound gas falls back towards the BH, shocks and accretion onto the BH produce a luminous flare that peaks in the UV/X-ray \citep{Lacy:1982,Rees:1988, Ulmer:1999}.  A number of analytic \citep[e.g.,][]{Phinney:1989, Kochanek:1994, Khokhlov:1996, Strubbe:2009} and numerical \citep[e.g.,][]{Evans:1989,Laguna:1993,Khokhlov:1993,Bogdanovic:2004} studies have explored the behavior of the debris after a MS star is tidally disrupted and the feasibility of using the luminosity and evolution of the flare to search for and determine properties of the BH.  Others have proposed flares and possible detonation of the star due to tidal compression as an additional observable consequence of tidal disruptions \citep{Carter:1982,Carter:1983, Bicknell:1983, Luminet:1985, Kobayashi:2004}.  Recent work along these lines suggests that the compression will produce shocks capable of raising the star's surface temperature to that of its core, resulting in a prompt flare \citep{Guillochon:2009, Brassart:2008, Brassart:2010}.  

Detailed models have also been employed to predict the rate at which MS stars are tidally disrupted by supermassive BHs.  However, uncertainties in the supermassive BH mass spectrum, the stellar distribution function in galactic bulges, and the relation between the two complicate calculations of this rate.   For example, improved measurements of the mean ratio of BH mass to bulge mass made between the studies conducted by \citet{Magorrian:1999} and \citet{Wang:2004} resulted in a factor of ten increase in the predicted event rate, raising it to $\sim 10^{-5}~\rm{yr^{-1}~Mpc^{-3}}$.  This rate is constrained by systematic surveys for tidal disruption flares.  Historically, around ten candidate tidal disruptions were first identified by serendipitous observations of UV/X-ray flares in surveys \citep[e.g.,][]{Grupe:1995,Brandt:1995,Komossa:1999,Komossa:1999a, Gezari:2006,Cappelluti:2009,Maksym:2010}.  Follow up observations of the variability of these sources ruled out other explanations \correct{and confirmed that the luminosity of the sources declined with the same $t^{-5/3}$ time dependance that models predict for the decline in the mass fallback rate \citep{Halpern:2004,Esquej:2008}}.  Subsequent studies have discovered additional tidal disruption candidates and, by accounting for the efficiency and sky coverage of the surveys, determined \correct{that} supermassive BHs tidally disrupt MS stars at a rate consistent with $10^{-5}-10^{-4}~\rm{yr^{-1}~ galaxy^{-1}}$ \citep{Donley:2002,Gezari:2008,Gezari:2009,Esquej:2008, Luo:2008}. The high energy accretion flares are the brightest and most immediate sign of a MS star being tidally disrupted by a BH, but they are not the only sign.    

In addition to the UV/X-ray emission,  \citet{Roos:1992} proposed that the unbound portion of the debris can be photoionized and produce optical emission lines.  The nature of the optical emission lines was explored in detail in the numerical models of \citet{Bogdanovic:2004} and the analytic models developed by \citet[hereafter SQ09]{Strubbe:2009}.  In each of these studies, the source of the ionizing photons is the accretion flare.  The prospect of optical emission from tidal disruptions is exciting for the following reasons.  First, SQ09 showed that tidal disruptions are bright enough in the optical band that they are potential sources for large transient surveys (e.g. The Large Synoptic Survey Telescope, The Palomar Transient Factory, and Pan-STARS), which suggests that the number of detected candidate events will increase in the near future.  Second, the emission-line signature of such an event helps us identify its nature and distinguish it from other, more common transients that can occur near the nuclei of galaxies, such as supernovae. 

While flares from tidally disrupted MS stars might be a more frequent probe of dormant BHs in galactic \correct{nuclei}, \citet{Sigurdsson:1997} pointed out that the BH can capture compact objects as well because this region is also populated with the products of stellar evolution.  In particular, \correct{white dwarfs (WDs)} are also subject to the processes that drive MS stars into orbits around the central BH with pericenter distances $R_{p} < R_{T}$, and can therefore be tidally disrupted as well.  However, since the mass of a WD is comparable to that of a low mass MS star and its radius is smaller by a factor of $\sim100$, $R_{T}$ is much closer to the BH for a WD than for a MS star.  Consequently, if $M_{BH} \ga 5 \times 10^{5} \msun$, $R_{T}$ for a WD is inside of the BH's Schwarzschild radius, $R_{S} = 2GM_{BH}/c^{2}$. When $R_{T} \le R_{S}$, the WD plunges into the BH horizon before being tidally disrupted, making such events undetectable by electromagnetic observations.  This makes WD tidal disruptions an excellent indicator of the presence of an IMBH \correct{in the center of a galaxy or globular cluster}, because if the central BH were more massive than $\sim 10^{5} \msun$ we would not observe anything.  

Like the MS case, using WD tidal disruption events to probe IMBH requires prior theoretical studies to determine how the observed flares are related to the properties of the WD-IMBH system.  While this scenario has not received as much attention as the MS case, the early stages of the tidal disruption of a WD in an unbound orbit around an IMBH were studied numerically by \citet{Frolov:1994}, who found that material flowed away from the WD in supersonic jets as it was disrupted.  \citet{Rosswog:2009} followed the evolution of the debris further and found that gas returns to the BH at a rate that is sufficient to power an accretion flare at the Eddington luminosity, $L_{Edd} = 1.3 \times 10^{41} (M_{BH}/10^{3}M_{\sun})\; \rm{erg\;s^{-1}}$, for up to one year.  Additionally, since it is supported by degeneracy pressure, when $R_{p} \la 0.5 R_{T}$ the tidal compression of the WD can lead to explosive nuclear burning \citep{Luminet:1989,Rosswog:2009}.  The latter authors found that, in the most favorable cases, the energy generated by thermonuclear burning during the compression phase exceeds the binding energy of the WD and is similar to that of a Type Ia supernova.  However, if the WD does not penetrate too deeply within $R_{T}$, the case is similar to that of the tidally disrupted MS stars described above, in which a portion of the debris is accreted by the BH and the rest of the material remains unbound and flows away from the BH.  \citet{Sesana:2008} calculated the light curves of optical and near-UV emission lines produced when the unbound debris is photoionized by the accretion flare and proposed these lines might be an electromagnetic counterpart to the gravitational radiation emitted as \correct{a WD initially on a bound orbit} spirals into the IMBH.  Here, we revise and update these results by \correct{considering the case of a WD initially on an unbound orbit and} using a more sophisticated model for the behavior of the debris.    

We apply the analytic description developed by SQ09 to study optical and UV emission lines from tidally disrupted MS stars to the case of a WD being tidally disrupted by an IMBH.  We discuss the evolution of the debris in \autoref{model}, drawing primarily from published work.  In \autoref{emission}, we describe photoionization models and the predicted emission lines from these events, including emission-line profiles.  Finally, we discuss our results in the context of resent observations in \autoref{discussion}.

\section{The Dynamical Model for the Debris}
\label{model}
We adopted the analytic prescriptions of SQ09 to model the evolution of the debris after a WD is tidally disrupted by an IMBH.  Their model describes a MS star approaching a supermassive BH on a parabolic orbit with pericenter distance $R_p \leq R_T$.  However, since all of the relevant quantities depend only on the BH mass, the pericenter distance, and the mass and radius of the star, the model can also be applied to the case of a WD being tidally disrupted by an IMBH.   Furthermore, even though the WD equation of state is stiffer than that of a MS star, the deviations in the tidal disruption process and the behavior of the debris that arise from this difference are small, so the approximate formulae developed in SQ09 give a reasonable description of the WD case.  The model assumes that half of the material becomes bound to the BH and forms an accretion disk after the WD is disrupted, and that the other half of the debris is ejected on hyperbolic orbits.  The details of these components in the SQ09 model and the adaptations we have made are described below.  For a full derivation of the model, we refer the reader to SQ09.  

The evolution of the debris was modeled for a range of initial conditions.  To explore how the optical emission lines are affected by the mass of the IMBH, we computed models with $M_{BH} = 10^{2}, 10^{3},\;\rm{and}\;10^{4}\msun$.  Even though a WD of the mass used in the models would be tidally disrupted outside of the event horizon of a $10^{5}\msun$ IMBH, \correct{we do not consider this case because the tidal disruption radius} is within the last stable orbit around the IMBH.  We also constructed models for two different values of the pericenter distance, one set with $R_{p} = R_{T}$ and, for comparison, a second set with $R_{p} = 0.3R_{T}$, even though \cite{Rosswog:2009} report that the WD would explode in the latter case.  \autoref{tab:ics} lists the initial conditions for each model we considered.  In each of the models, we set $M_{WD} = 0.55\msun$ because the mass distribution of WDs observed in the Sloan Digital Sky Survey is strongly peaked at this value \citep{Madej:2004}.  We calculated the WD's radius \correct{to be $R_{WD} = 8.6 \times 10^{8}$ cm} using the mass-radius relation of \citet{Nauenberg:1972}.  

\subsection{The Accretion Disk}
\label{disk}
As the bound material falls back to pericenter, it shocks on itself and the orbits of the debris circularize to form an accretion disk.  In the model, the disk extends from the last stable orbit, $R_{LSO}$, out to approximately $2 R_p$.  Like SQ09, we considered both Schwarzschild BHs with $R_{LSO} = 3 R_S$ and maximally spinning BHs with $R_{LSO} = R_S/2$.  \correct{The rate at which the bound material falls back towards the BH is set by the time it takes the debris to return to pericenter and the mass of the WD, and was computed using equations (1) and (2) of SQ09.  The mass fallback rate decreases as $t^{-5/3}$, and at early times it can exceed the Eddington mass accretion rate,  $\dot{M}_{Edd} = 10 L_{Edd}/c^2$.  We capped the rate at which material is accreted by the BH, $\dot{M}$, at $\dot{M}_{Edd}$ and, following SQ09, we assumed that the excess mass is blown off in a radiation driven wind.  Because the nature of these outflows is uncertain,} we did not consider emission from the outflows \correct{or how the outflows might affect the emission lines produced in the unbound material.  SQ09 suggested that the material in the outflows can also be photoionized and produce additional line luminosity, making the predictions of this model lower limits.  \correct{In a followup paper, \citet{Strubbe:2010} consider the photoionization of the out-flowing gas that is outside the photosphere of the outflow and calculate the resulting {\it absorption-line} spectrum.} Furthermore, the outflows could alter the radiation field that photoionizes the unbound material and, thus, change the ionization state of the unbound debris.  Since these effects were not accounted for in our model, the emission line luminosities predicted during the super-Eddington phase are uncertain.  The outflows subside after a time $t_{\rm Edd}$, given in equation (3) of SQ09 as $t_{Edd} \sim 0.1\, (M_{BH}/10^{6}\msun)^{2/5}\,(R_{p}/3 R_{S})^{6/5}\, (M_{\star}/\msun)^{3/5}\, (R_{\star}/R_{\sun})^{-3/5}$ yr. In \autoref{tab:ics} we list the value of $t_{Edd}$ for each of our models.} 
    
\subsection{The Unbound Debris}
\label{fan}
\begin{figure}
	\centering
	\includegraphics[width=1\textwidth]{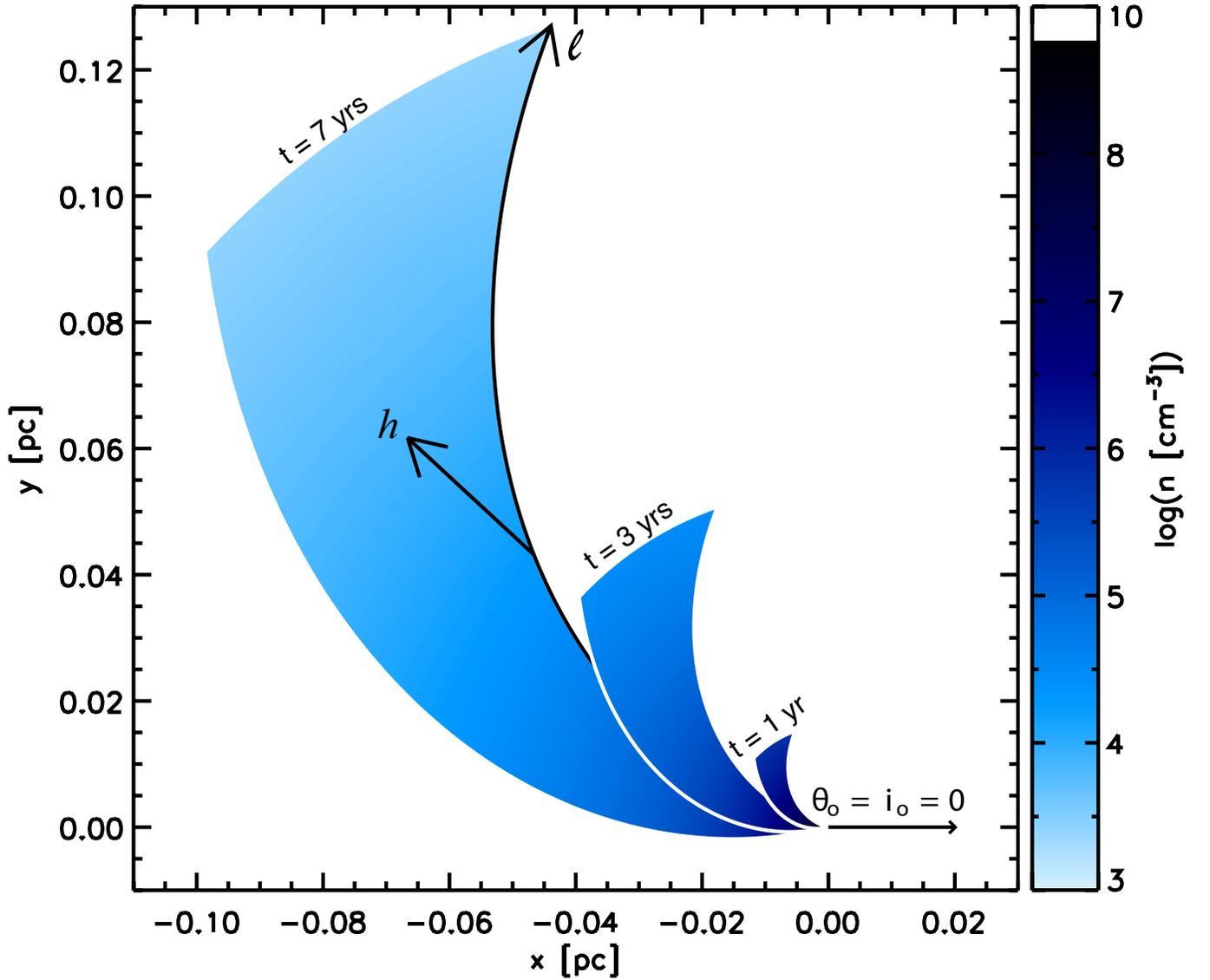}
	\caption{The spatial distribution of the unbound debris from model C, projected into the orbital plane. The BH sits at the origin and the WD approached it from the $-x$ direction on a parabolic orbit.  The debris is shown one, three, and seven years after the tidal disruption.  The shading shows the density of the debris, which goes at $R^{-3}$.  The $\ell$-curve begins at the BH and lies along the illuminated face of the debris cloud.  The ionizing radiation from the disk encounters the cloud at a different angle of incidence for each point along $\ell$ and travels along the corresponding $\vec{h}$ into the cloud.  The length of $\vec{h}$ shown here is greatly exaggerated, the incident radiation only reaches depths $\la 10^{-3} R$. \label{fig:fan}}
\end{figure}

As the star approaches the BH, it is tidally stretched along the direction of its motion and then spun up, resulting in a spread in the specific energy of the debris \citep{Rees:1988}.  SQ09 calculated how this range of specific energies leads to an expanding, curved wedge of unbound debris spanning a range of azimuthal angles and radial distances from the BH (see \autoref{fig:fan}).  The accretion flare illuminates the inside face of this wedge, and the gas along this edge reprocesses the radiation and produces the optical emission lines we are interested in.  SQ09 modeled the extent of this face in terms of $R_{\max}, \Delta\phi$, and $\Delta i$, the maximum radial distance from the BH, the azimuthal dispersion of the debris in the orbital plane, and the angular dispersion of the debris \correct{perpendicular} to the orbital plane, respectively.  We adopt their equations for \correct{the angular dispersion of the unbound material, however, we make a minor modification to calculate the distance between the BH and material along the debris arc.}  Since we considered several points along the arc, not just the most distant material, we calculated the distance to the BH at any point along the arc with
\begin{equation}
	R(\phi) \sim v_{p}\,t\cot\left(\frac{\phi}{2}\right),
	\label{eqn:r}
\end{equation}
where $v_{p}$ is the pericenter velocity of the WD, $t$ is the time since \correct{pericenter passage}, and $\phi$ is the angle in the orbital plane between pericenter and the point in the debris cloud being considered and is in the range $(\pi - \Delta\phi) < \phi < \pi$.  \correct{The WD is initially on a parabolic orbit, so $v_{p} = (2GM_{BH}/R_{p})^{1/2}$.}  SQ09 calculated $R_{\max}$ in a similar manner and our equation \eqref{eqn:r} is consistent with the derivation of \citet{Evans:1989} for the values of $\phi$ considered here.  \correct{In \autoref{tab:ics}, we list $\Delta\phi$ and $v_{\rm max}/c$, the maximum velocity in the unbound debris cloud as a fraction of the speed of light, for each set of initial conditions.} In addition to $R$ and $\phi$, we define the $\ell$\correct{-curve}, which begins at the IMBH and lies along the illuminated face of the debris cloud.

Furthermore, because we computed photoionization models at several points along the inside \correct{face} of the cloud of unbound debris, we also had to deviate from SQ09's model when we calculated the density of the unbound debris, $n$.  \citet{Kochanek:1994} found that the density in the stream of unbound material was not uniform and decreased with distance from the BH as $n \propto R^{-3}$ \correct{because the material is undergoing free expansion}.  We used this density profile in our model and normalized it using the fact that the mass of the wedge of unbound material was half the mass of the WD.

\section{Predicted Emission}
\label{emission}
\begin{figure}
	\centering
	\includegraphics[width=0.6\textwidth,angle=90]{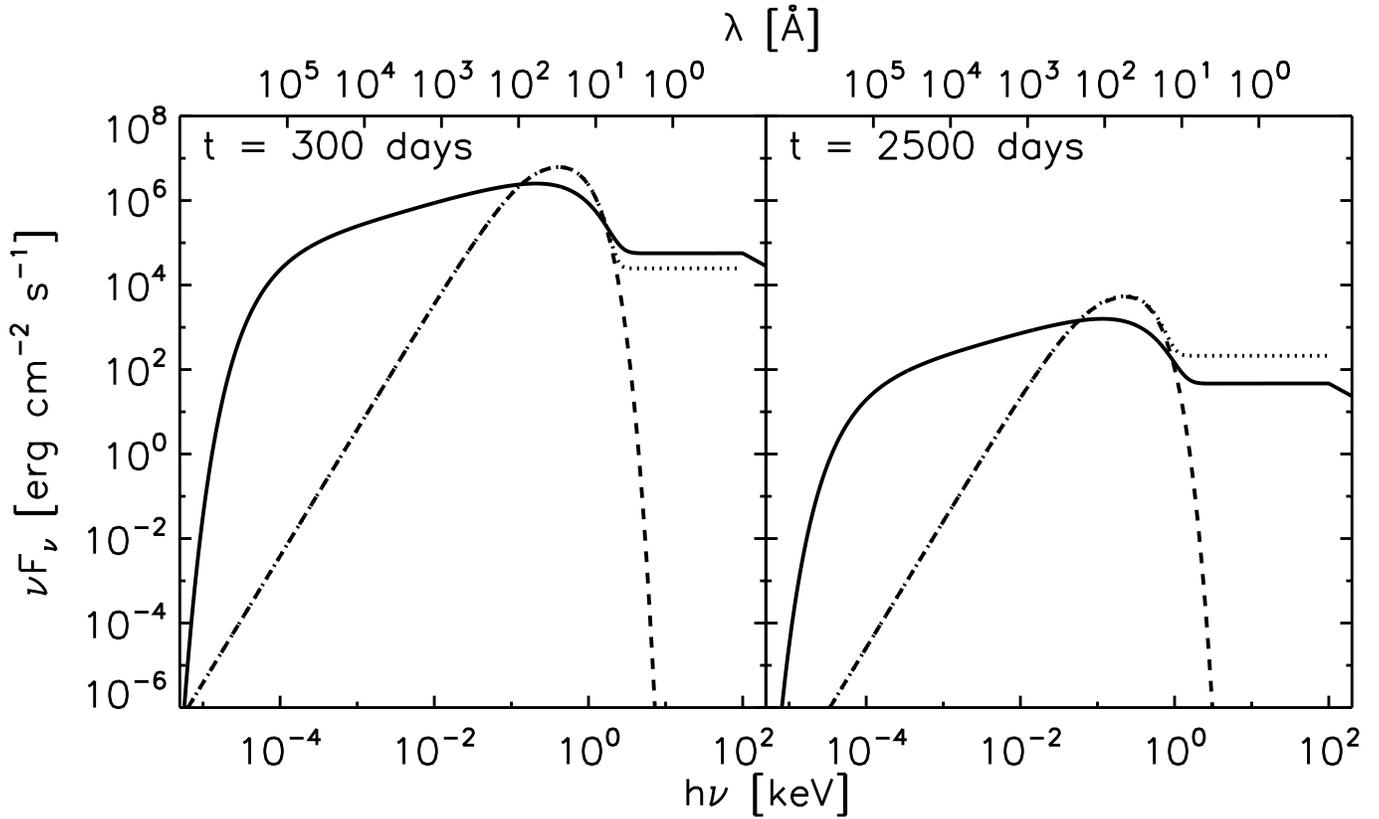}
	\caption{The parameterized AGN (solid line), multicolor blackbody (dashed line), and the multicolor blackbody plus X-ray power law (dotted line) SEDs incident on an azimuthal segment of the debris cloud considered in model C.  The segment has $R =1.9 \times 10^{16}$ cm and $R = 1.6\times 10^{17}$ cm at $t = 300$ days and $t = 2500$ days, respectively.  Note that over time the position of the peak moves towards longer wavelengths as the temperature in the disk decreases and the total flux decreases as the accretion rate declines and the segment moves further away from the ionizing source.\label{fig:seds}}
\end{figure}

Predicting the emission lines we will observe when a WD is tidally disrupted by an IMBH requires a model for the emission from both the bound and unbound portions of the debris.  To calculate the luminosity and spectrum of the accretion flare generated by the disk of bound gas, SQ09 assumed  a ``slim disk'' model \citep{Abramowicz:1988} and derived an expression for the effective \correct{temperatures as a function of radius in} the disk (see their equation (19)).  Then, using the range of temperatures throughout the disk, they modeled the spectrum as a multicolor blackbody.  We considered this spectral energy distribution (SED) for the disk as well as two others.  We also used a SED that consisted of the multicolor blackbody and an additional X-ray component from $0.1 - 100$ keV with a power law of the form $f_{E} \propto E^{-1}$, \correct{where $f_{E}$ is the flux density per unit energy and $E$ is the photon energy.}  We normalized the intensity of the X-ray power law using the intensity of the multicolor blackbody at $2500$ \AA\; and the $\alpha_{OX}$ parameter\footnote{The optical-to-X-ray spectral index is the exponent of a power-law connecting the monochromatic luminosity densities at 2500~\AA\ and 2~keV. If $L_{\nu}\propto\nu^{-\aox}$, then $\aox \equiv $\hfill\break $- \left[\log L_{\nu}(2500\;{\rm \AA}) - \log L_{\nu}(2~{\rm keV})\right] / \left[\log \nu(2500\;{\rm \AA}) - \log \nu(2~{\rm keV})\right] = 1+0.384\; \left[\log{(\nu L_{\nu})_{\rm 2500\;\AA} -\log(\nu L_{\nu})_{\rm 2\; keV}}\right]$.} based on the findings of  \citet{Steffen:2006}.  The third SED we considered was an empirically constructed, multi-component model for the SED of an active galactic nucleus (AGN) that was first used in \citet{Korista:1997} and explored in greater detail by \citet{Casebeer:2006}.  This model is parameterized by the temperature of the peak of the model's thermal component, $\alpha_{OX}$, and the slope of the the X-ray component.  We set these parameters to the temperature of the peak of the multicolor blackbody, $-1.4$, and $-1$, respectively.  \correct{The accretion disk-IMBH system is similar to an AGN, so we used these three SEDs to sample the range of UV/X-ray and UV/infrared ratios observed in AGN.  This is necessary because heating of the free electrons in the unbound material by infrared radiation can change its ionization structure and irradiating this material with X-rays can lead to significant heating per ionization, which would affect the luminosities of collisionally excited lines.}  In all three cases, the time evolution of the SED is governed by the declining mass accretion rate, which decreases the \correct{bolometric luminosity} of the SED and causes its peak to shift to longer wavelengths as time goes on.  \autoref{fig:seds} shows each SED and its evolution.  These SEDs were used as one of the inputs to the photoionization models described in the next section.     

\subsection{Photoionization Models}
\label{photoion}
The unbound material is assumed to radiate because it is illuminated by the accretion flare described above.  Predicting the emission lines generated by this material requires photoionization models to determine how the gas reprocesses the incident radiation.  We performed the photoionization calculations with version 08.00 of Cloudy, last described by  \citet{Ferland:1998}.  Since the debris cloud consists of material from a tidally disrupted WD, its composition is different from that of a MS star.  The composition of WDs is predicted by the yields of nuclear reactions and confirmed with asteroseismology \citep{Metcalfe:2003}.  Following \cite{Madej:2004}, we assumed mass fractions of 67\% O, 32\% C, and 1\% He and all other elements with  their relative abundances scaled from the solar values.  Hydrogen is assumed to make up 0.001\% of the WD's mass.  For each set of initial conditions (i.e., models A-F in \autoref{tab:ics}), we calculated the state of the accretion flare SED and the wedge of unbound debris at a series of times after the tidal disruption.  For each time step, we split the wedge into six azimuthal segments.  \correct{With six segments, we were able smoothly interpolate quantities along the $\ell$ direction.}  Next, we calculated the density of the gas and the intensity of the ionizing radiation at the center of each segment.  Several considerations went into the calculation of the latter.  First, each segment is at a different distance from the accretion disk and has a correspondingly different geometric attenuation factor.  Next, since the segments lie along an arc, the ionizing radiation's angle of incidence is also different for each segment.  Finally, in addition to changes in the intensity of the incident radiation, the shape of the SED incident on each segment also varies.  The light travel time, $t_{lt}$, to the unbound debris furthest from the BH is comparable to the timescale on which the accretion flare evolves, so the SED of the radiation incident on these portions of the wedge was emitted at $t-t_{lt}$, when the peak was at a higher energy.  \correct{Each of these considerations has a significant effect on our models.  Geometric attenuation and light travel time dominate for more distant regions, but the material nearest the disk is illuminated at nearly glancing incidence, significantly reducing the intensity of the incident radiation despite the proximity to the disk.}  With these parameters, we are able to calculate how the unbound debris responds to the ionizing radiation emitted by the accretion disk.

\begin{figure}
	\centering
	\includegraphics[width=1\textwidth]{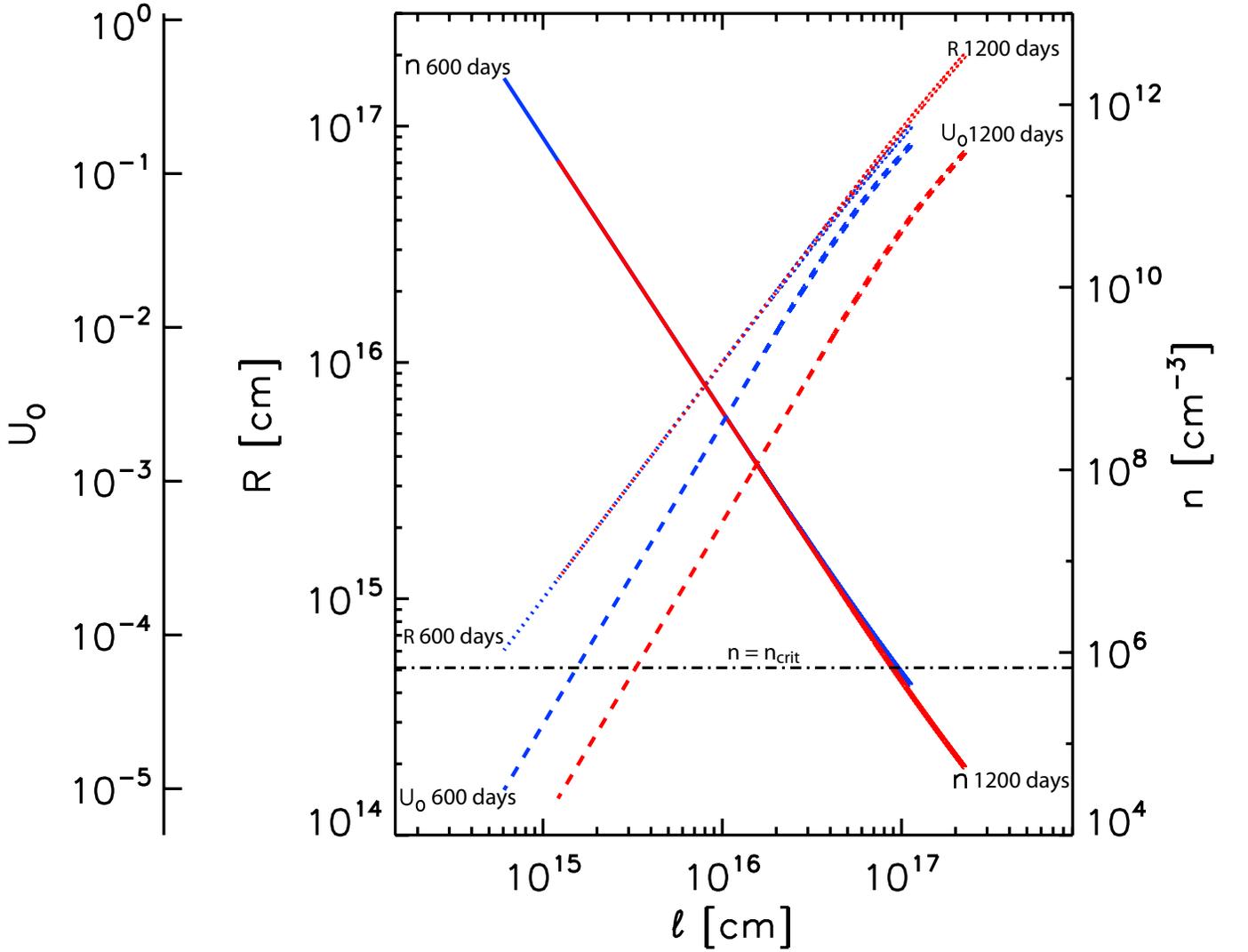}
	\caption{The oxygen ionization parameter $U_{O}$ (dashed line), radial distance from the IMBH $R$ (dotted line), and the number density $n$ (solid line) along $\ell$, the illuminated face of the cloud (see \autoref{fig:fan}). These quantities are plotted at 600 (blue) and 1200 (red) days after the tidal disruption for model C.  At these times, the super-Eddington outflows have subsided so $U_{O}$ declines as $t^{-0.2}$.  Note that $R$ is nearly proportional to $\ell$.  The horizontal, dot-dashed line shows the critical density of [\ion{O}{3}] $\lambda5007$.  At later times, a larger portion of the cloud is below $n_{cr}$ and the region of maximum emissivity, the portion of the cloud with $n = n_{cr}$, shifts to a part of the cloud moving with a lower velocity with respect to the IMBH.\label{fig:l} }
\end{figure}

The ionization state of the gas is conventionally described by the ionization parameter $U_{H} = Q_{H}/(4\pi n_{H}R^{2}c)$, where $Q_{H}$ is the emission rate of photons with energy greater than 1 Ry, $n_{H}$ is the hydrogen number density, and $R$ is the distance to the ionizing source.  However, since hydrogen is depleted in the material considered here, it is advantageous to use a more abundant reference element when calculating the ionization parameter.  We chose to use oxygen because its first ionization potential is very near that of hydrogen so the ratio of the two ionization parameters is $U_{O}/U_{H} = X_{H}/X_{O}$, where $X_{O}$ and $X_{H}$ are the oxygen and hydrogen abundances by number, respectively.  \correct{Thus, $U_{O}$ is the ratio of the density of photons capable of singly ionizing oxygen to the total density of oxygen atoms.}  \autoref{fig:l} shows how $U_{O}$, the total number density \correct{of all elements}, and the radial distance to the IMBH change along the illuminated face of the debris cloud, $\ell$, for model C at 600 and 1200 days after tidal disruption.  Since the density drops faster with distance from the accretion disk than the flux of ionizing photons does, $U_{O}$ increases along the $\ell$-curve.  \correct{While the fallback rate is super-Eddington}, the value of $U_{O}$ increases linearly with time at all points along the \correct{$\ell$-curve} because the luminosity of the ionizing source remains constant at $L_{Edd}$, while the density of the expanding debris cloud decreases.  \correct{We find numerically that $U_{O}$ declines approximately as $t^{-0.2}$ after the super-Eddington accretion phase ends\footnote{\correct{This agrees with a simple analytic estimate,  $U_{O}$ is the ratio of the density of ionizing photons to the density of oxygen atoms so $U_{O} \propto L_{d}/(T_{d} R^{2}n) \propto t^{-1/4}$ where $L_{d}$ and $T_{d}$ are the bolometric luminosity and maximum temperature in the accretion disk, respectively.  The time dependance is slightly steeper than in our models because this estimate does not adequately account for the broad peak in the SEDs.}}.}   

\begin{figure}
	\centering
	\includegraphics[width=0.6\textwidth,angle=90]{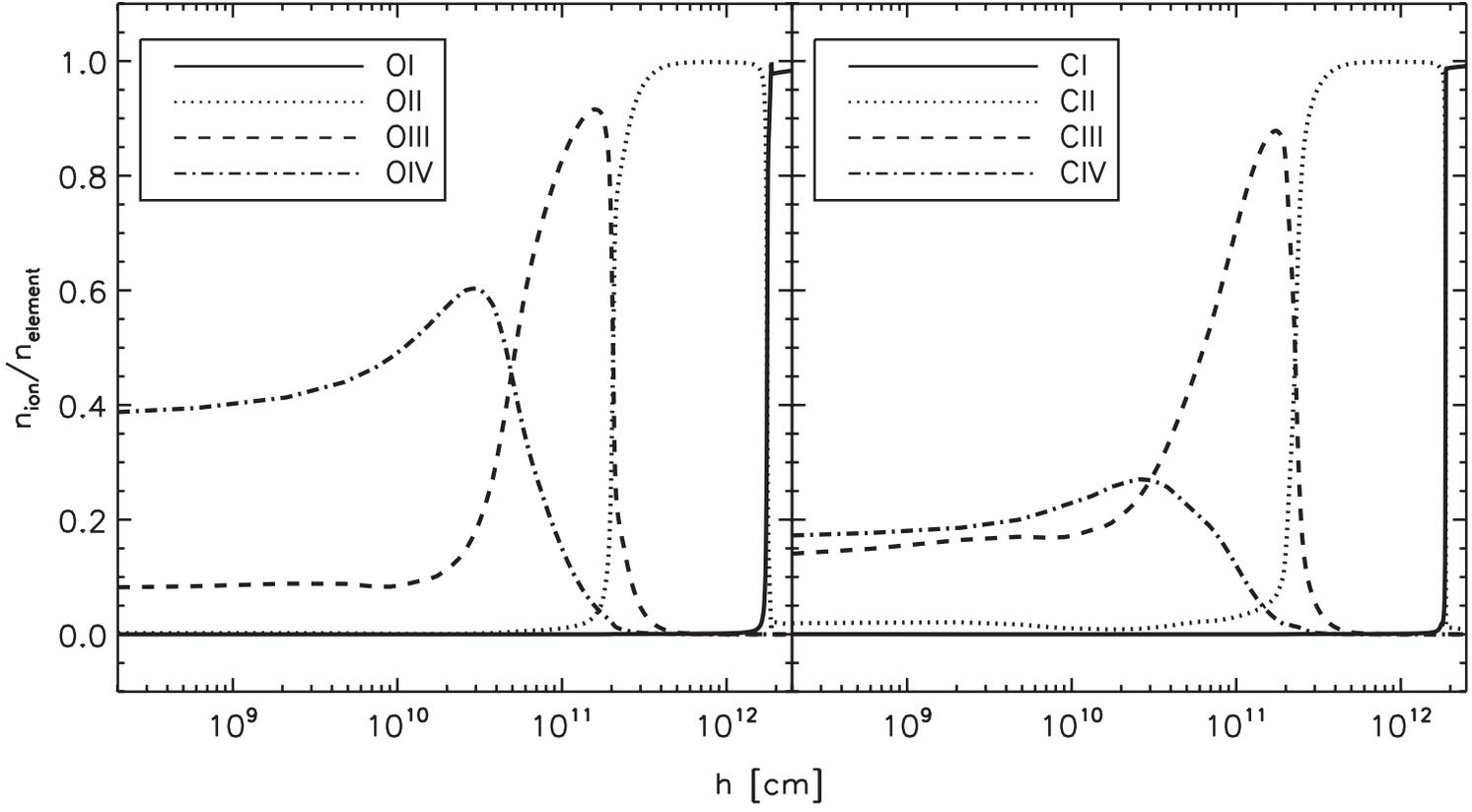}
	\caption{The fractions of the first four ions of oxygen (left panel) and carbon (right panel) along $\vec{h}$.  These values are from a photoionization calculation for an azimuthal segment near the middle of the debris arc with $R = 1.8 \times 10^{17}$ cm, $n = 7.5 \times 10^{4}~{\rm cm^{-3}}$, and $U_{O} = 0.08$, 1500 days after tidal disruption in model C, using the multicolor blackbody plus X-ray power law SED.  Highly ionized species are present in the surface layers of the cloud and singly ionized carbon and oxygen are the dominant ions in deeper layers. In most of our photoionization models the dominant ions of oxygen and carbon are \ion{O}{2}  and \ion{C}{2}.\label{fig:fracs}}          
\end{figure}

\correct{The fractions of the first four ions of oxygen and carbon along $\vec{h}$ from a representative photoionization calculation are shown in \autoref{fig:fracs}.  In nearly all of our photoionization models with $R_{p} = R_{T}$ (i.e., models A,C,E, and F), the dominant ion of oxygen is \ion{O}{2} and the dominant ion of carbon is \ion{C}{2}.  The exceptions are the azimuthal segment nearest the IMBH at times less than 300 days after tidal disruption, which consists mostly of neutral oxygen and carbon, and the segment farthest from the IMBH at times greater than 2500 days after tidal disruption, whose dominant ion are \ion{O}{3} and \ion{C}{3}.  As the density and ionization parameter change, the relative thickness of the zone dominated by each ion of oxygen does not change significantly between 300 and 2500 days after tidal disruption, so the relative abundances of the ions of oxygen do not vary significantly in these models.  In the case of carbon, however, the abundance of \ion{C}{3} relative to \ion{C}{4} decreases during the first 600 days after tidal disruption, before leveling off.  During this time some of the \ion{C}{3} is photoionized into \ion{C}{4}.  For the models with $R_{p} = 0.3 R_{T}$ (i.e., models B and D), \ion{O}{2} and \ion{C}{2} are the dominant ions of oxygen and carbon, respectively, throughout the debris tail for the first 300 days.  In these models, the density of the unbound debris is much lower, so $U_{O}$ is much higher.  Therefore, at later times, higher ionization species (\ion{O}{4}, \ion{O}{5}, \ion{C}{4}, and \ion{C}{5}) dominate in different portions of the cloud.}

In the photoionization models, we make the simplifying assumption that the density of the debris is constant along $\vec{h}$, the path the ionizing radiation follows into the debris cloud (see \autoref{fig:fan}).  This assumption is valid because the thickness of the ionized layer is very small compared to the distance to the BH, so the density in this skin only deviates from the surface value by 0.1\% in the most extreme cases.  In addition, we assumed that the ionized gas remains in photoionization equilibrium during the 4000 days probed by our models.  This requires that the density of the gas and flux of ionizing radiation change on timescales greater than the recombination time $t_{rec} \sim (n_{e}\alpha_{rec})^{-1}$ of the ions of interest.  Since the cloud consists mostly of carbon and oxygen, there are many electrons per ion, which keeps the electron density reasonably large even at late times when the density of the debris cloud is low.  Furthermore, the range of conditions encountered in the models yielded \ion{O}{4}, \ion{O}{3}, \ion{O}{2}, \ion{C}{4}, and \ion{C}{3} recombination coefficients that, in combination with the electron density, kept the recombination time well below the timescales on which the conditions in the wedge of unbound debris changed \correct{(the ranges of recombination rates are $6.7 \times 10^{-7} - 1.9 {\rm~s^{-1}}$, $1.3 \times 10^{-7} - 0.52{\rm~s^{-1}}$, $4.4 \times 10^{-8} - 0.15 {\rm~s^{-1}}$, $1.3 \times 10^{-7} - 0.97 {\rm~s^{-1}}$, and $3.8 \times 10^{-7} - 0.98 {\rm~s^{-1}}$ for the above ions at the lowest and highest electron densities, respectively). }  Therefore, the cloud can remain in photoionization equilibrium for over ten years.

\subsection{Line Profiles}
\label{profiles}
Combining the dynamical and photoionization models allows us to calculate the observed profile of the \correct{emission lines in the} [\ion{O}{3}] $\lambda\lambda$4959, 5007 doublet.  The dynamical model we have adopted gives us an analytic expression for the velocity of the material along the illuminated edge of the debris cloud ($\ell$) in the orbital plane, $v_{R}(\phi) \sim v_{p} \cot(\phi/2)$.  This is the dominant component of the debris cloud's velocity, so we neglected the motion of the material perpendicular to the orbital plane, which is smaller by a factor of $R_{WD}/R_{p}$ (SQ09).  The velocity was projected onto the observer's line of sight, which is defined by $i_{0}$ and $\theta_{0}$, the angles between the line of sight and the orbital plane and the pericenter direction, respectively.  \correct{\autoref{fig:fan} shows the $i_{0}=\theta_{0} = 0$ axis.  We interpolated the [\ion{O}{3}] $\lambda5007$ flux computed in our photoionization models along the $\ell$-curve to determine the restframe flux at each velocity along $\ell$.  We took the flux at 4959 \AA~ to be one third of this value.}  We considered Doppler boosting of this emissivity, but the effect is insignificant at most points along the arc and produces a maximal change of $1.2 \%$.  Finally, at each point, we took into account the relativistic Doppler shift, the gravitational redshift, and local broadening.  \correct{We found that the gravitational redshift was negligible in all of our models.}  The local broadening was assumed to \correct{have a Gaussian profile} and included terms for the thermal motions of the gas, the range of velocities within each bin along the \correct{$\ell$-curve}, and the range of velocities in the ionized skin, along $\vec{h}$.  \correct{Since the conditions along the $\ell$-curve varied significantly, each source of local broadening considered in our models dominates in a different portion of the illuminated face of the cloud at any given time.  Furthermore, the conditions in the unbound debris change substantially over time, so the dominant source of local broadening in a given portion of the debris cloud also changes over time.} In the next section, we discuss the results of the photoionization and emission-line profile calculations.

\subsection{Results}
\label{results}
\begin{figure}
	\centering
	\includegraphics[width=1\textwidth]{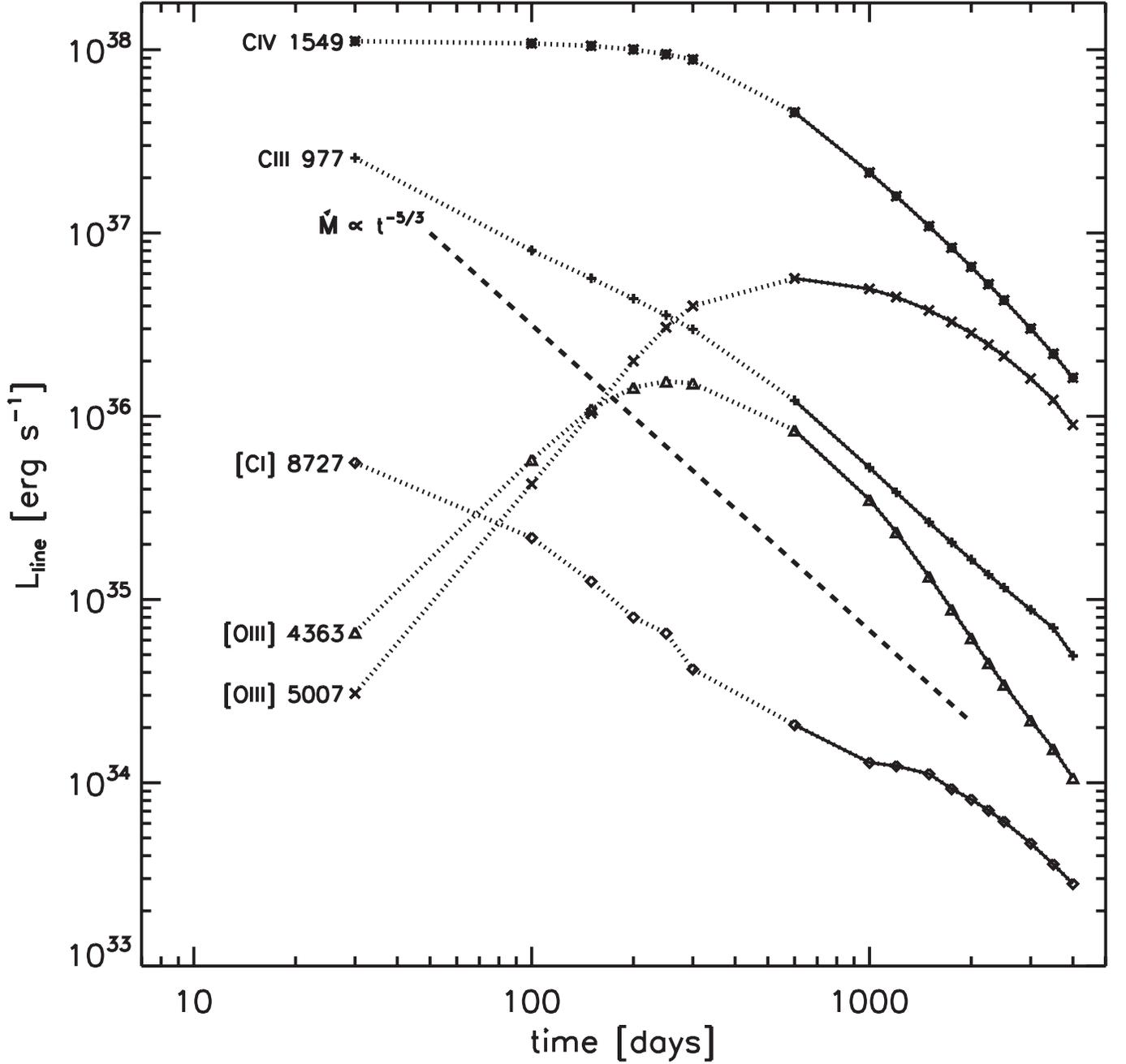}
	\caption{The light curves of five of the strongest emission lines.  These light curves come from dynamical model C and were calculated using the multicolor blackbody plus X-ray power law SED. The dotted portion of each curve shows the line luminosity while the fallback rate is super-Eddington and our model is uncertain.  The dashed line illustrates the $t^{-5/3}$ dependance of the mass fallback rate.  The normalization of the dashed line is arbitrary and is shown for comparison with the slopes of the emission line light curves.     \label{fig:lums}}
\end{figure}

Our photoionization calculations predict an emission-line spectrum that is dominated by lines from carbon and oxygen.  The light curves of some of the strongest emission lines are shown in \autoref{fig:lums} and the luminosities of the six strongest features at two different times are listed in \autoref{tab:lum}.  The values given in \autoref{fig:lums} and \autoref{tab:lum} are from calculations preformed with the multicolor blackbody plus X-ray power law SED, but calculations with the multicolor blackbody SED produced similar results \correct{because the ionization state of the gas is largely determined by the flux of UV photons, which is nearly identical in these two SEDs.}  In the case of the multicomponent, empirical AGN SED, the emission-line luminosities we calculated were consistent with those calculated using the other two SEDs, only varying by a factor of $\sim 2$ in the most extreme cases.  The AGN SED has much stronger optical and near-IR continuum emission, which \correct{allows the continuum to outshine the emission lines, greatly reducing their equivalent widths.  However, these results indicate that heating of the free electrons in the unbound material by IR radiation from the IMBH-accretion disk system does not have a strong effect on the emission line luminosities.  Thus the main difference between the results from the AGN SED and the other two cases is that the optical continuum is higher and equivalent widths of the optical emission lines are lower, even though their luminosities are approximately the same.}  Our discussion will focus on the spectra predicted by calculations using the other two SEDs, which are nearly identical in the optical and UV bands.  

For each model with $R_{p} = R_{T}$ (i.e., models A, C, E, and F), the UV lines \ion{C}{4} $\lambda$1549 and \ion{C}{3} $\lambda977$ have the highest luminosity early on.  The luminosities of these and other permitted lines decline over time.  \correct{The luminosity of the \ion{C}{3} $\lambda977$ line declines faster than the  \ion{C}{4} $\lambda$1549 line because of the reduction in the amount of \ion{C}{3} relative to \ion{C}{4} discussed in \S3.1.}   After 100 days, the [\ion{O}{3}] $\lambda5007$, [\ion{O}{3}] $\lambda4363$, and [\ion{O}{2}] $\lambda7325$ lines have the largest equivalent widths. \correct{These are forbidden transitions, which are collisionally excited and then de-excite radiatively as long as the electron density is below the threshold for collisional de-excitation (i.e., the critical density for the transition $n_{cr}$, listed in Table~\ref{tab:lum} for reference). In the early phases of the evolution of the debris (the super-Eddington phase), the luminosity of these forbidden lines rises with time and the lines reach their maximum luminosities at 600, 250, and 300 days, respectively, when the outer tip of the ionized debris tail is at the corresponding critical density. After the peak, the line luminosities begin to decline, like the permitted lines. To understand this trend, we note that, at densities below the critical density of a transition the line emissivity per unit volume increases linearly with density and peaks at the critical density. As the density increases above the critical density, the line emissivity per unit volume drops by a factor of several (with the depletion of the upper level population by collisions) and then levels off (as the upper level is re-populated by collisions and radiative decays from higher levels).\footnote{We verified this behavior, by solving the population equations for the [\ion{O}{3}] $\lambda5007$ transition and then combining the resulting level populations with the luminosity per unit density of ions in the lower level of the transition. To carry out this exercise, we assumed a 5-level ion and followed the methodology in \S3 of \citet{Osterbrock:2006}.} Thus, the initial rise in the luminosity of the forbidden lines is the result of the expansion of the debris while the density is above the critical density of the transition. This expansion occurs during the super-Eddington phase, when the ionizing luminosity is constant. During this expansion, the peak luminosity in the light curves corresponds to the time when the density at the outer tip of the debris tail reaches the critical density. After that time the density of an increasing portion of the debris drops below the critical density (causing a reduction in the emissivity per unit volume) and the total luminosity declines.  We can quantify the rate of the initial rise in a forbidden emission line's luminosity if it occurs while the ionizing luminosity is constant and the density in the debris cloud is above the critical density.  As the debris expands, the column density of the ionized skin remains constant because the decline in the flux of ionizing photons resulting from the increasing distance from the IMBH is balanced by the increase in the cloud's surface area.  This column sustains a constant emission line flux at the surface of the cloud, so the rise in the forbidden emission line luminosity is driven by the cloud's expanding emitting area.  This area grows as $t^{2}$, and this evolution can be seen in Figures \ref{fig:lums} and \ref{fig:oiii}.}

We show the [\ion{O}{3}] $\lambda$5007 and  \ion{C}{4} $\lambda$1549 light curves for each of the models in \autoref{fig:oiii}.  These curves come from calculations preformed with the multicolor blackbody plus X-ray power law SED and we note that calculations that exclude the X-ray component of the incident radiation predict luminosities within a few percent of those shown.  We calculated the luminosities by multiplying the emergent flux computed by Cloudy in each azimuthal segment by that segment's emitting area and then summing the luminosities across the entire illuminated face.  The light curves indicate how the evolution of the [\ion{O}{3}] $\lambda$5007 line depends on the properties of the WD-IMBH system and how its evolution can be used to identify such a tidal disruption event.  First, the luminosity of the line at any given time increases as the mass of the IMBH decreases.  This is because both the initial, maximal mass accretion rate and the emitting area of the cloud of unbound debris decrease as the mass of the IMBH increases.  Next, the line reaches its peak luminosity at earlier times for smaller values of $R_{p}/R_{T}$ \correct{(i.e., models B and D).}  This is primarily because the spread in the debris is larger for smaller $R_{p}$ and the critical density of the transition is reached earlier.  \correct{The steeper decline of the [\ion{O}{3}] $\lambda$5007 luminosity in models B and D is also a result of the lower density in these debris tails.  The less dense debris becomes highly ionized, reducing the amount of \ion{O}{3} in the cloud.}   In any case, the decline of the \correct{line} luminosity after it reaches its maximum value is less step than the $t^{-5/3}$ decline of the mass fallback rate.  Finally, if the IMBH is spinning, the inner radius of the accretion disk is much closer to the IMBH and therefore has a \correct{maximum temperature that is 2.7 times higher than that of a disk around a Schwarzschild BH}.  This leads to an increase in the flux of ionizing photons and a corresponding overall increase in the luminosity of the [\ion{O}{3}] $\lambda$5007 line.  

The emission-line profile of the [\ion{O}{3}] $\lambda\lambda$4959, 5007 doublet depends on the properties of the WD-IMBH system.  Qualitatively, the full width at half maximum (FWHM) of each line is relatively small shortly after the tidal disruption.  At these times, \correct{the debris cloud is above the critical density and the line profile is dominated by emission from the outer tip of the debris cloud where the emitting area is maximal.  The spread of velocities in this small segment of the cloud is comparatively narrow, resulting in an emission line with a small FWHM.  This material has the greatest velocity with respect to the IMBH, so the peak of the emission line will be offset from its rest value.  Eventually, the density of the material at the outer tip of the debris tail drops to $n_{cr}$ and the emission line reaches its maximum luminosity.  As time goes on, material at smaller $\ell$, that is moving at slower velocities relative to the IMBH, reaches $n_{cr}$ for the [\ion{O}{3}] $\lambda\lambda$4959, 5007 doublet and the emissivity per unit volume increases. At the same time, the increasing volume of material in the more distant regions of the cloud partially compensates for the decreasing emissivity per unit volume of this material.  As a result, material at a large range of velocities produces the [\ion{O}{3}] $\lambda\lambda$4959, 5007 doublet at comparable luminosities and the FWHM of each line increases.  } The broadening is asymmetric because the velocity of the gas along the illuminated edge increases monotonically with distance from the IMBH.  The orientation of the observer determines the degree of asymmetry and whether the \correct{wing blueward or redward of the peak is broader.}  Furthermore, the position of the peak of each line also shifts as the debris cloud expands and \correct{the region of maximum emissivity moves toward the IMBH along the $\ell$-curve.  The emissivity is largest when $n = n_{cr}$, and as the cloud expands, regions near the IMBH traveling at lower velocities relative to the IMBH reach this density and become brighter at 5007 \AA.  Later, when most of the unbound debris is below the critical density, the brightest region can shift away from the IMBH towards a region with a larger volume, and in some cases a larger fraction of \ion{O}{3}.}  Again, whether the peak shifts towards the red or the blue is determined by the orientation of the observer.  The range of velocities along the inner edge of the debris cloud is broader for higher mass BHs\footnote{\correct{SQ09 showed that the maximum velocity of the unbound material is $v_{\max} \propto v_{p}R_{p}^{-1/2}$.  For tidal disruption, $R_{p}\la R_{T}\propto M_{BH}^{1/3}$.   Since $v_{p}\propto (M_{BH}/R_{p})^{1/2}$, the maximum velocity of the unbound material increases as $M_{BH}^{1/6}$}} , but since the FWHM of observed line profiles depends on the velocity along the line of sight, broader lines do not necessarily mean more massive BHs.  

\begin{figure}
	\centering
	\includegraphics[width=0.5\textwidth]{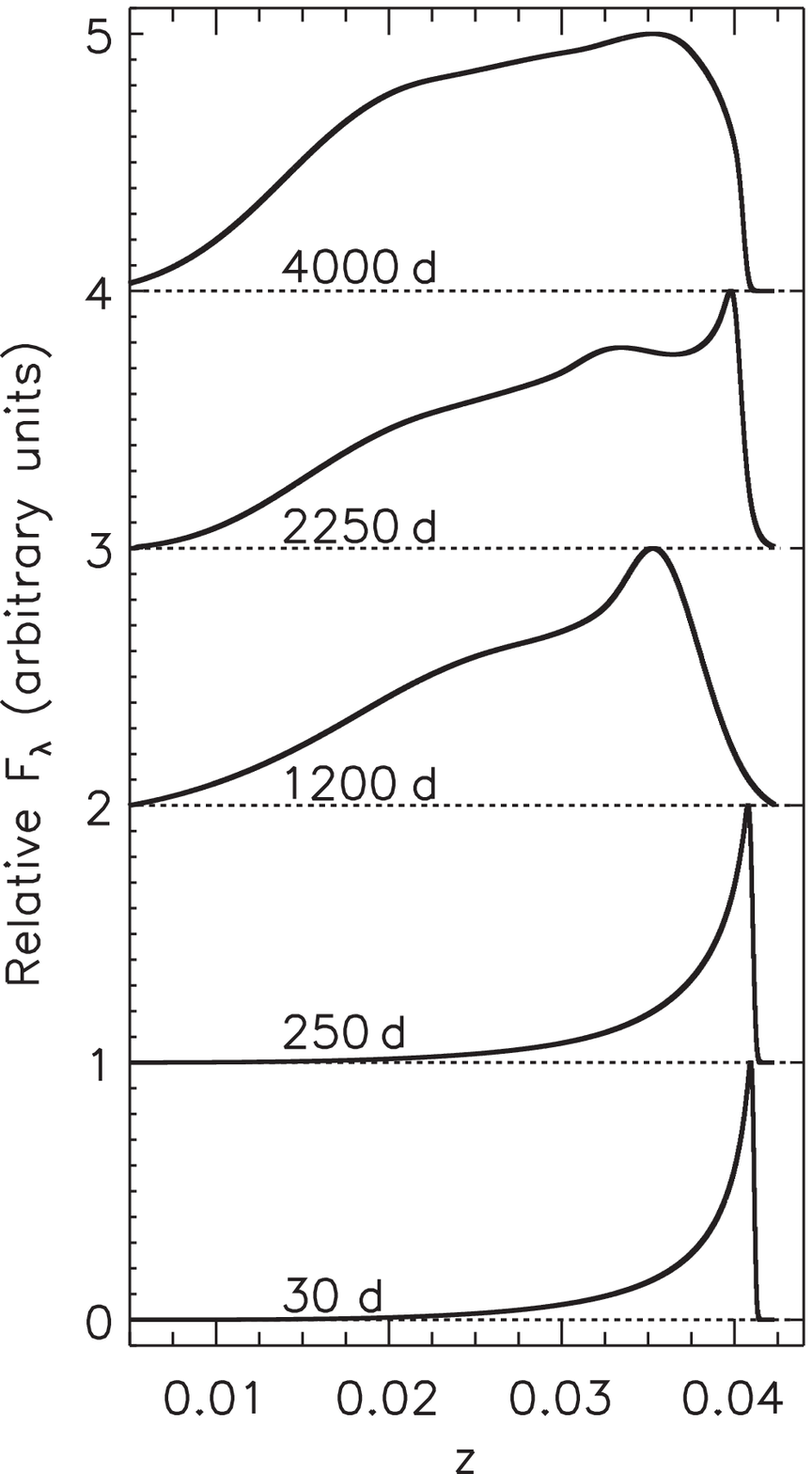}
	\caption{The time evolution of the [\ion{O}{3}] $\lambda5007$ emission-line profile as seen by an observer at $i_{0} = 45\degr$ and $\theta_{0} = -5\degr$.  Time increases from bottom to top and the time for each profile is written on the plot.  The dotted lines show the $F_{\lambda} = 0$ level for each profile.  The horizontal axis shows $z = (\lambda - 5007{\rm\AA})/5007{\rm\AA}$.  Each profile has been normalized so that it has a maximum height of 1 and the profiles are offset by one unit for clarity.  Note that in addition to becoming broader with time, the peak of the line profile shifts to lower velocity.  Once most of the cloud is below $n_{cr}$, the emissivity per unit volume decreases throughout the cloud and the peak shifts back towards higher velocities that correspond to regions in the tail with larger volume and a larger fraction of \ion{O}{3}.   \label{fig:broad}}
\end{figure}

\begin{figure}
	\centering
	\includegraphics[width=0.5\textwidth]{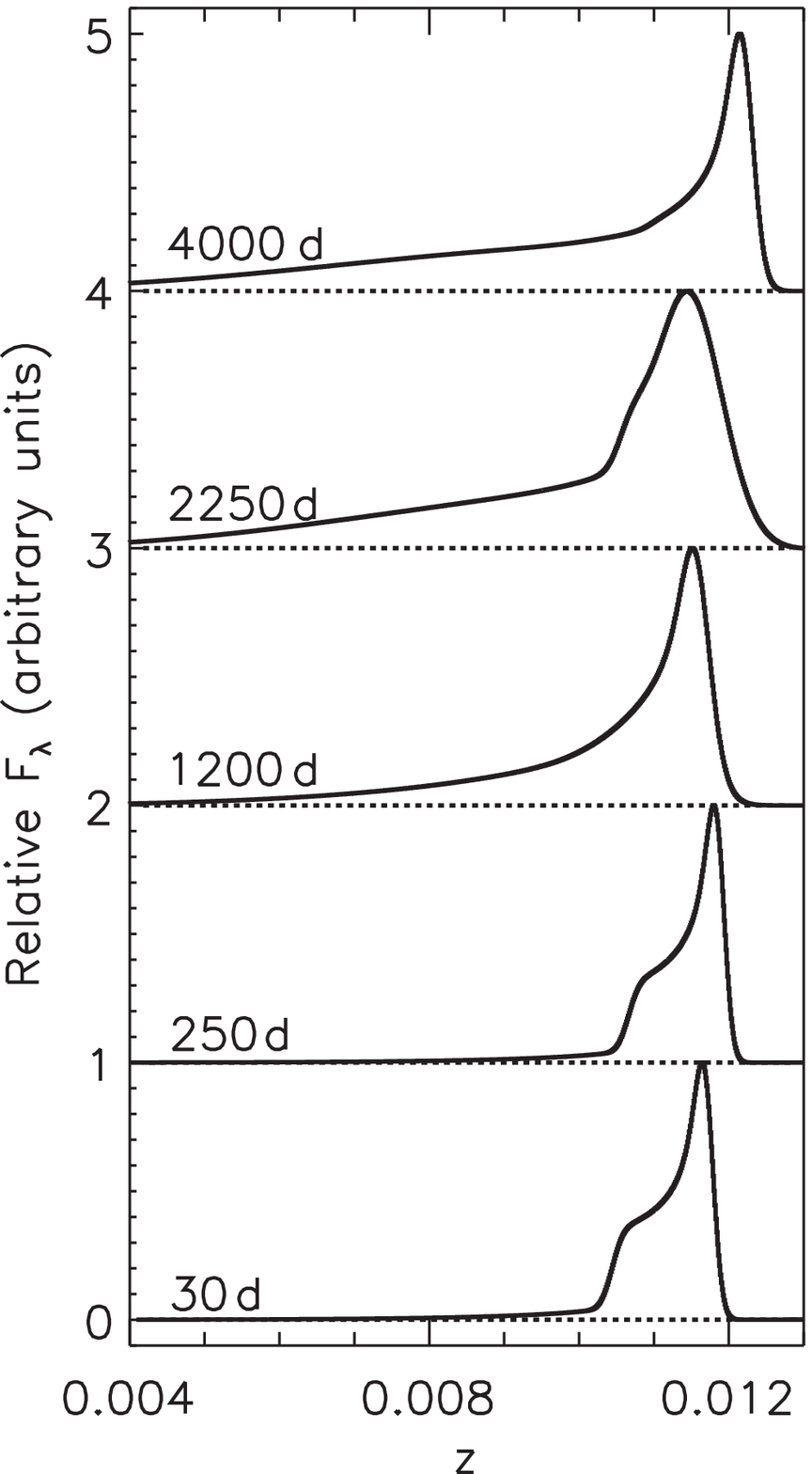}
	\caption{The time evolution of the [\ion{O}{3}] $\lambda5007$ emission-line profile as seen by an observer at $i_{0} = 45\degr$ and $\theta_{0} = 30\degr$. Time increases from bottom to top and the time for each profile is written on the plot.  The dotted lines show the $F_{\lambda} = 0$ level for each profile.  The horizontal axis shows $z = (\lambda - 5007{\rm\AA})/5007{\rm\AA}$.  Each profile has been normalized so that it has a maximum height of 1 and the profiles are offset by one unit for clarity.  Note that in addition to becoming broader with time, the peak of the line profile shifts to lower velocity. Once most of the cloud is below $n_{cr}$, the emissivity per unit volume decreases throughout the cloud and the peak shifts back towards higher velocities that correspond to regions in the tail with larger volume and a larger fraction of \ion{O}{3}.   \label{fig:shoulder}}
\end{figure}

To illustrate these points, Figures \ref{fig:broad} and \ref{fig:shoulder} show the time evolution of the [\ion{O}{3}] $\lambda5007$ line for model C, as observed from $\theta_{0} = -5\degr$ and $\theta_{0} = 30\degr$, respectively, \correct{with $i_{0} = 0$ in both figures.}  In the former, the observer is positioned such that the spread in line of sight velocities along the brightest part of the cloud is quite large.  In the latter, the velocity of the material is mostly transverse to the line of sight, so the range of observed velocities in the brightest part of the cloud is relatively small.  \correct{These figures do not include the [\ion{O}{3}] $\lambda$4959 line, so the distinct ``components'' in the line profile result from the distribution of [\ion{O}{3}] $\lambda5007$ emitted flux as a function of projected velocity.}  Unfortunately, since changes in the position of the observer result in such drastic changes in the line profile, it is difficult to extract detailed information about the nature of the WD-IMBH system from it.  However, as we discuss in the following section, the general properties of the emission-line profiles are useful in identifying candidate WD tidal disruptions.            

\section{Discussion}
\label{discussion}
\begin{figure}
	\centering
	\includegraphics[width=0.5\textwidth,angle=90]{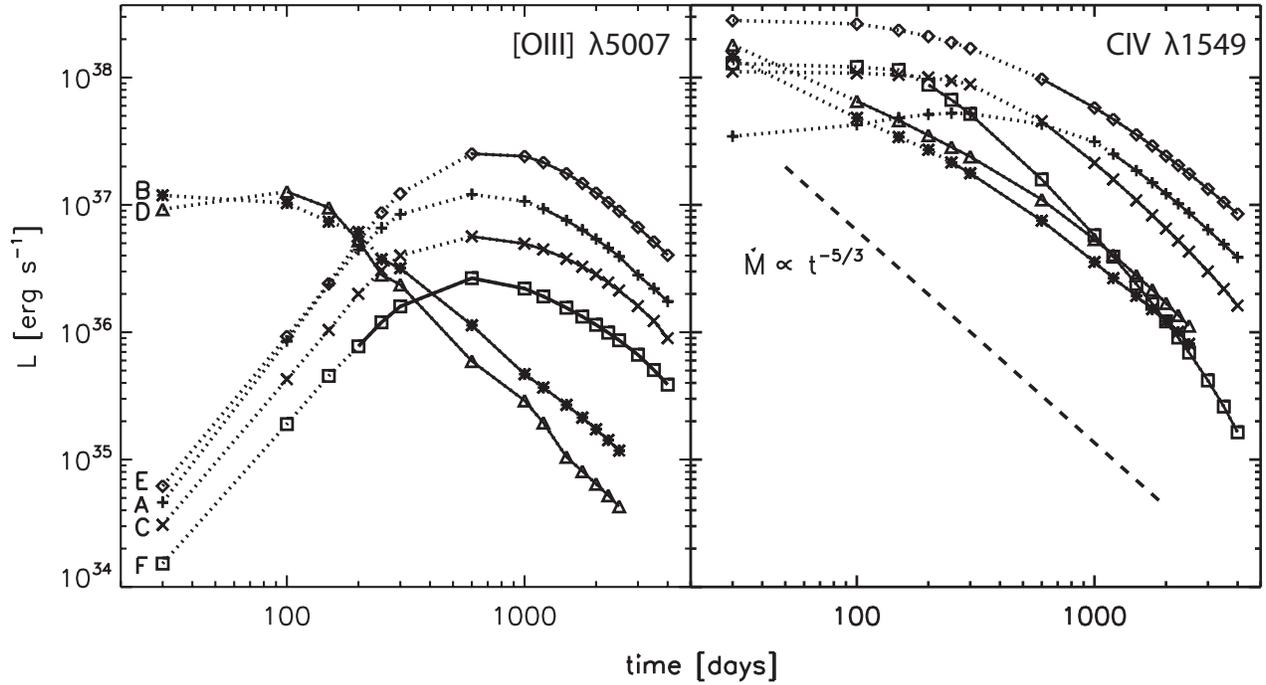}
	\caption{The [\ion{O}{3}] $\lambda5007$ (left panel) and \ion{C}{4} $\lambda1549$ (right panel) light curves for each model given in \autoref{tab:ics}.  The photoionization calculations that produced these light curves used the multicolor blackbody plus X-ray power law SED; calculations that exclude the X-ray component produce nearly identical results. The dotted portion of each curve shows the line luminosity while the fallback rate is super-Eddington and our model is uncertain. The curves in the left panel differ from those in the right because [\ion{O}{3}] $\lambda5007$ is a forbidden emission line and the density in the debris cloud must drop to the critical density for this transition for this emission line to form.  Curves B and D in the left panel are different from the others because they correspond to models with $R_{p} = 0.3 R_{T}$.  Similarly, the curves in the right panel corresponding to models with $R_{p} = 0.3 R_{T}$ differ from all the others by their steep decline at early times.  The dashed line illustrates the $t^{-5/3}$ dependance of the mass fallback rate.  The normalization of the dashed line is arbitrary and is shown for comparison with the slopes of the emission line light curves.  \label{fig:oiii}  }
\end{figure}

We have presented calculations of the line emission produced when the unbound debris of a WD that has been tidally disrupted by an IMBH is illuminated by the associated accretion flare.  We found that the two strongest emission lines, by far, are \ion{C}{4} $\lambda1549$ \correct{at early times} and [\ion{O}{3}] $\lambda5007$ \correct{at late times}.  Furthermore, the emission lines formed in the illuminated face of the debris cloud will have broad, asymmetric profiles with velocity widths of several hundred to a few thousand $\rm{km~s^{-1}}$.  Our results suggest that UV and optical spectra whose strongest features are broad  \ion{C}{4} $\lambda1549$ and [\ion{O}{3}] $\lambda5007$ emission lines can serve to identify WD tidal disruption events, when they are detected in addition to an X-ray flare in the center of a globular cluster or dwarf elliptical galaxy.  \autoref{fig:oiii} shows how the characteristic timescales and luminosities of these emission lines scale with $R_{p}$ and the mass of the IMBH.  The luminosities of the \ion{C}{4} $\lambda1549$ and [\ion{O}{3}] $\lambda5007$ lines decline more slowly if $R_{p} = R_{T}$ than they do if $R_{p} = 0.3 R_{T}$.  For a constant $R_{p}$, the shapes of the emission-line light curves do not change much when the mass of the IMBH is changed, however, systems with lower mass IMBHs produce brighter \ion{C}{4} $\lambda1549$ and [\ion{O}{3}] $\lambda5007$ emission.  \correct{The emission line luminosities are sensitive to the density of the photoionized debris, and since this density is only weakly dependent on $M_{BH}$, the line luminosities evolve on similar timescales despite a factor of 100 increase in $M_{BH}$.  In these models $n = n_{0}/R^{3}$, where the normalization factor $n_{0} = M_{WD}^{5/6}\;M_{BH}^{1/6}\;(8\mu)^{{-1}}\,(R_{p}/3R_{T})^{1/2}\;\{\tan[(\pi-\Delta\phi)/2] [\sin\Delta i\,(\log(\sin(\Delta\phi/2))+6)]^{-1}\}$ ensures that the mass of the unbound debris tail is $0.5M_{WD}$.  Here we have used the mean molecular mass $\mu$.  Numerical fits to the quantity in the \correct{curly} brackets show that it increases with the mass of the IMBH as $M_{BH}^{0.54}$ when $R_{p}/R_{T}$ is held constant.  Altogether, we find that the density of the unbound debris is only weakly dependent on the mass of the IMBH, $n\propto M_{BH}^{0.21}$, so it is difficult to constrain $M_{BH}$ with the [\ion{O}{3}] $\lambda5007$ light curve.}       

Our work drew heavily on the SQ09 model for the tidal disruption of a MS star by an supermassive BH, so we summarize the uncertainties in the model described there.  First, since the accretion disk becomes geometrically thin after $\dot{M}$ drops below the Eddington rate, it becomes susceptible to viscous instabilities that impede the steady flow of material through the disk and this could result in a deviation from the smooth $t^{-5/3}$ dependance of the ionizing radiation.  \correct{Material will accumulate in the disk rather than accreting onto the BH when disk's viscous timescale becomes comparable to the time since tidal disruption.  This occurs at $t\sim 8 - 12$ yr for the disks in our models.}  Furthermore, the model is no longer valid when the mass accretion rate drops below $\sim 0.01 \dot{M}_{Edd}$ and the disk becomes radiatively inefficient.  The final time steps in our models B-F are nearing this limit, and this change will lead to a reduction in the number of UV photons not accounted for in this model.  \correct{Since the photoionization cross-sections for O and C decline significantly from UV to X-ray energies, the ionization state of the unbound debris depends sensitively on the flux of UV photons whose energies are near the ionization potentials of C and O.}  Next,  a significant amount of initially bound material can be driven away from the IMBH in the super-Eddington outflows.  \correct{The impact that the outflows have on emission from the unbound material is uncertain.  SQ09 pointed out that the outflows could also be photoionized by the accretion flare and generate additional line luminosity.  Additionally, since we did not account for how the outflows effect the radiation that illuminates the unbound debris, the emission line luminosities predicted for $t < t_{\rm Edd}$ are uncertain.}     

The modifications we made to the SQ09 model allowed us to test how some of their simplifications affect the predictions of the model.  In their photoionization calculations, they simplify the geometry of the wedge of unbound debris by approximating it as a cloud separated from the ionizing source by $R_{\max}$ (i.e. the maximum of equation \eqref{eqn:r} or $R(\pi - \Delta\phi)$).  Furthermore, they assume that the density of the unbound material is uniform and evaluate it by placing half the mass of the star in a region of volume $R_{\max}^{2}\,\Delta R\, \Delta i\, \Delta \phi$.  In our calculations, on the other hand, we have accounted for the changing ionization state of the material along the illuminated edge of the cloud by allowing the density and intensity of the ionizing radiation to vary along this edge, as discussed in \S 3.1.  Both methods produce similar emission-line light curves, with the SQ09 model predicting slightly higher luminosities for most lines.  This was true for both the WD-IMBH case and the MS-supermassive BH case.  Our models also show that the choice of SED for the accretion flare does not have a strong affect on the emission line luminosities.  We conclude, therefore, that their simpler method is sufficient to predict the luminosities and light curves of optical emission lines from tidally disrupted stars.

One additional source of uncertainty in our models is the composition of the unbound material.  The models of \citet{Luminet:1989} and  \citet{Rosswog:2009} found that compression of the WD as it passes through pericenter resulted in nuclear burning.  The amount of energy released by these reactions and how significantly they alter the composition of the debris depends on the initial composition of the WD, $M_{BH}$, and $R_{p}$.  To explore how a change in the abundances brought about by nuclear burning would influence the emission lines emitted by the unbound debris, we also computed photoionization models using abundances given in \citet{Rosswog:2009} for a $1.2\msun$ WD \correct{($R_{WD} = 2.7 \times 10^{8}$ cm)} that is tidally disrupted by a $500 \msun$ IMBH with $R_{p} = R_{T}/3.2$.  For the first year, the brightest lines are the UV lines \ion{C}{4} $\lambda1549$, \ion{Si}{3} $\lambda1888$, and \ion{O}{6} $\lambda1035$.  Later however, the emission-line spectrum is dominated by forbidden iron lines rather than carbon and oxygen lines.  The most luminous lines include [\ion{Fe}{10}] $\lambda6373$,  [\ion{Fe}{7}] $\lambda6807$, [\ion{Si}{8}] $\lambda1446$, [\ion{Fe}{7}] $\lambda3759$, [\ion{Fe}{5}] $\lambda3892$, and [\ion{Fe}{7} ]$\lambda5271$.  While, this scenario represents an extreme case in which the energy released by nuclear burning is greater than the binding energy of the WD, the presence of these iron and silicon lines in a spectrum that is otherwise similar to that described above could be evidence of nuclear burning during the tidal disruption.      

The next question to consider is how often we should expect to see emission lines from tidally disrupted WDs.  \citet{Ramirez-Ruiz:2009} estimate that the rate of $L_{Edd}$ flares from MS stars being tidally disrupted by $10^{3}-10^{4} M_{\sun}$ IMBHs in globular clusters is $\sim 4000~ \rm{yr^{-1}~ Gpc^{-3}}$.  Following these authors, we assume all globular clusters have an IMBH in their center and adopt a globular cluster space density of $n_{GC} \sim 4~\rm{Mpc}^{-3}$ \citep{Brodie:2006}.  We take the WD tidal disruption rate of $10^{-8}~\rm{yr^{-1}~(globular~cluster)^{-1}}$ from \cite{Sigurdsson:1997} and arrive at a WD tidal disruption rate of $\sim 40~ \rm{yr^{-1}~ Gpc^{-3}}$, 100 times lower than the MS rate.  This rate is a very optimistic estimate because it is unlikely that every globular cluster has an IMBH in its center and the actual rate could be \correct{much} lower.  The likelihood of observing such an event is further reduced by its relatively short lifetime of $\sim 10 {\rm~yr}$.   

Despite the meager rate and limited observing window,  \citet{Irwin:2010} proposed the tidal disruption of a WD by an IMBH as an explanation for two recent observations of X-ray and optical emission in the centers of globular clusters.  \citet{Irwin:2010} report on the ultraluminous X-ray source CXOJ033831.8-352604.  The source has a $0.3-10$ keV luminosity of $\sim2\times10^{39}~\rm{erg~s^{-1}}$ as well as [\ion{O}{3}] $\lambda$5007 and [\ion{N}{2}] $\lambda$6583 emission lines with luminosities of few $\times~10^{36}~\rm{erg~s^{-1}}$ and FWHMs of 140 $\rm{km~s^{-1}}$.  The authors discuss a second, similar source in the center of a globular cluster associated with NGC 4472.  \cite{Maccarone:2007} conclude that the source's X-ray luminosity of $4 \times 10^{39} ~\rm{erg~s^{-1}}$ and its variability can only be explained by the presence of a BH.  In follow up optical observations, \cite{Zepf:2008} found that the [\ion{O}{3}] $\lambda5007$ line luminosity was $1.4 \times 10^{37} ~\rm{erg~s^{-1}}$ and that the line's FWHM was 1500  $\rm{km~s^{-1}}$.    

\begin{figure}
	\centering
	\includegraphics[width=0.5\textwidth]{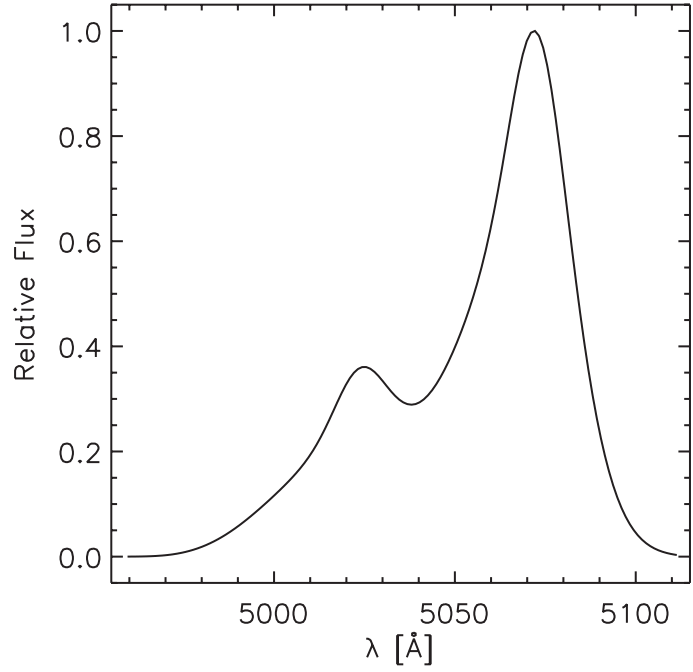}
	\caption{The [\ion{O}{3}] $\lambda\lambda$4959, 5007 line profile for model C, 1750 days after tidal disruption, as seen by an observer at $i_{0} = 89\degr$ and $\theta_{0} = -12\degr$.  The profile has been smoothed to a spectral resolution of 400 for comparison with observations of \citet{Zepf:2008}.  The profile is normalized to have a height of 1 and does not include contributions from the continuum.}
	\label{fig:profile}
\end{figure}

In both cases, the X-ray luminosity of the source is consistent with the models \correct{that assumed a multicolor blackbody plus X-ray power law SED for the accretion flare}.  However, in all of our models, the luminosity of the [\ion{N}{2}] $\lambda$6583 line is two orders of magnitude lower than that of the [\ion{O}{3}] $\lambda$5007 line.  Furthermore, the FWHMs of our synthesized emission-line profiles are significantly broader than 140 $\rm{km~s^{-1}}$.  These two discrepancies disfavor the interpretation of CXOJ033831.8-352604 as the tidal disruption of a WD by and IMBH.   The observations of optical emission lines by \citet{Zepf:2008}, however, are consistent with the model in terms of both luminosity and FWHM.  The observed luminosity of the [\ion{O}{3}] $\lambda$5007 line falls within the range of luminosities predicted for systems with IMBH masses between 100 - 1000 $\msun$.  \autoref{fig:profile} shows a synthesized [\ion{O}{3}] $\lambda\lambda$4959, 5007 emission-line profile that has been smoothed to the same spectral resolution as the observations of \citet{Zepf:2008}.  The line profile was synthesized for model C with the observer at $i_{0} = 89\degr$ and $\theta_{0} = -12\degr$.  \correct{We have chosen these values to match the shape of the observed emission-line profile.}  The first similarity between the model and the observations is that the lines are so broad that they are blended.  The FWHM of the model line profile is $1700~\rm{km~s^{-1}}$.  Also, both the observed and modeled profiles are asymmetric, with the blue wing much broader than the red.  \correct{This suggests that we are oriented with the flow such that $v_{\rm max}$ is nearly parallel to our line of sight.} While the synthesized and observed profiles are qualitatively similar, spectra of high signal-to-noise ratio and higher spectral resolution are required for a more detailed comparison to determine whether or not the observed line profiles are consistent with the model.

One major drawback to this interpretation of the data involves the duration of the X-ray flare and the observed  [\ion{O}{3}] $\lambda5007$ luminosity.  An X-ray luminosity of $8.5 \times 10^{39} {\rm~erg~s^{-1}}$ was measured for this source with {\it ROSAT} in 1992 \citep{Colbert:2002}.  The change in luminosity between the {\it ROSAT} observation and the observation reported in \cite{Maccarone:2007} is consistent with the model for the X-ray flare, if the {\it ROSAT} observation detected the flare $\sim 1$ year after the WD was tidally disrupted.  However, this means that  the [\ion{O}{3}] $\lambda$5007 luminosity was still $1.4 \times 10^{37} ~\rm{erg~s^{-1}}$ more than 15 years after the tidal disruption, and long after the peak in [\ion{O}{3}] $\lambda$5007 emission predicted by this model (see \autoref{fig:oiii}).  The model can only explain this [\ion{O}{3}] $\lambda$5007 luminosity if the IMBH in this globular cluster is $100 \msun$ and maximally spinning.  Even though it is possible to tune the model to explain the observed X-ray and [\ion{O}{3}] $\lambda$5007 luminosities, because the high emission line luminosity persists for many years longer than it would in all but the most favorable scenario, it is unlikely that this source is a WD that has been tidally disrupted by a IMBH, according to the model.  \correct{Furthermore, the large amplitude X-ray variability of the source reported in \citet{Maccarone:2010} is also not consistent with the model and \citet{Steele:2010} find that the profile of the emission lines in the [\ion{O}{3}] $\lambda\lambda$4959, 5007 doublet are well fit by a two component model with a bipolar conical outflow and a lower velocity Gaussian component.}   

If, in spite of the above discrepancies, this source is a WD that has been tidally disrupted by a IMBH, then the strongest line in a UV spectrum will be \ion{C}{4} $\lambda1549$.  Unlike the [\ion{O}{3}] $\lambda5007$ emission line, the \ion{C}{4} $\lambda1549$ flux is fairly constant across the illuminated face of the cloud so most of this line's luminosity is generated in the outer portions of the debris tail where the emitting area is largest.  This results in a emission-line profile consisting of a redshifted core with a FWHM of several hundred $~\rm{km~s^{-1}}$ and a broad blue shoulder. Furthermore, if the X-rays decline dramatically in the future, the [\ion{O}{3}] luminosity should also dim over the course of one year according to this model.  The delay in the decline is due to the light travel time from the IMBH to the distant portions of the unbound debris that generate most of the [\ion{O}{3}] luminosity.

Given the low rate at which WDs are tidally disrupted by IMBHs, the short time during which the disruption is observable, and the simplifying assumptions made in the model presented here, it is not surprising that the emission from CXOJ033831.8-352604 and the globular cluster associated with NGC 4472 are not completely consistent with the predictions of the model.  On average, detecting a single WD tidal disruption in a globular cluster requires surveying $10^{6}$ galaxies, so it is unlikely that one has been observed.  Additionally, improved models that consider the dynamics of the unbound debris in more detail and include emission from the material ejected in the super-Eddington outflows could capture behavior in the optical emission line spectrum that is not predicted by this model.  However, given the capabilities of current and planed optical and X-ray transient surveys, the number of candidate WD tidal disruption events should increase in the near future.  If follow up optical and UV spectroscopy do not reveal strong, broad  [\ion{O}{3}] $\lambda5007$ and  \ion{C}{4} $\lambda1549$ lines that evolve as shown if \autoref{fig:oiii}, it is unlikely that the emission is produced by a WD that has been tidally disrupted by an IMBH.

\acknowledgements

We thank the anonymous referee for many detailed and thoughtful comments.  We also thank Steve Zepf, Steinn Sigurdsson, Linda Strubbe, and Eliot Quataert for helpful discussions.  This work was supported in part by the Zaccheus Daniel Fellowship.

\bibliography{clausen}

\begin{deluxetable}{ccccccc}
\tablecolumns{7}
\tablewidth{0pc}
\tablecaption{Model Parameters \label{tab:ics}}
\tablehead{
\colhead{} & \colhead{$M_{BH}$} & \colhead{\underline{$R_{p}$}} & \colhead{\underline{$R_{LSO}$}} & \colhead{$t_{\rm Edd}$\tablenotemark{a}} & \colhead{$\Delta\phi$\tablenotemark{b}} & \colhead{\underline{$v_{max}$\tablenotemark{c}}}\\
\colhead{Model} & \colhead{($M_{\sun}$)} & \colhead{$R_{T}$} &\colhead{$R_{S}$} &\colhead{(days)} &\colhead{(rad)} & \colhead{$c$}}
\startdata
A&$10^{2}$&$1$&$3$&$1100$&$1.5$&0.032\\
B&$10^{2}$&$0.3$&$3$&$240$&$2.7$&0.083\\
C&$10^{3}$&$1$&$3$&$440$&$0.99$&0.052\\
D&$10^{3}$&$0.3$&$3$&$95$&$1.9$&0.15\\
E&$10^{3}$&$1$&$0.5$&$440$&$0.99$&0.052\\
F&$10^{4}$&$1$&$3$&$170$&$0.68$&0.082
\enddata
\tablenotetext{a}{The length of the super-Eddington mass fallback phase}
\tablenotetext{b}{The azimuthal dispersion of the unbound debris in the orbital plane}
\tablenotetext{c}{The maximum velocity in the unbound debris cloud as a fraction of the speed of light}
\end{deluxetable}

\begin{deluxetable}{ccccccc}
\tablecolumns{7}
\centering
\tablewidth{0pc}
\tablecaption{Predicted Line Luminosities \label{tab:lum}}
\tablehead{
\colhead{} & \colhead{Luminosity of}               & \multicolumn{5}{c}{Luminosities of Selected Lines Relative to [\ion{O}{3}]$\;\lambda$5007}  \\ 
\colhead{} & \colhead{[\ion{O}{3}]$\;\lambda$5007} & \multicolumn{5}{c}{\hrulefill}  \\ 
\colhead{Model} & \colhead{(erg~s$^{-1}$)} & \colhead{[\ion{O}{3}]$\;\lambda$4363} & \colhead{[\ion{O}{2}]$\;\lambda$7325} & \colhead{\ion{C}{3}$\;\lambda$977} & \colhead{\ion{C}{4}$\lambda$1549} & \colhead{[\ion{C}{1}]$\;\lambda$8727} \\ 
\colhead{$n_{\rm cr}$ (cm$^{-3}$)\tablenotemark{a}~$\rightarrow$} & \colhead{$7\times 10^5$} & \colhead{$3\times 10^7$} & \colhead{$6\times 10^6$} & \colhead{} & \colhead{} & \colhead{$\sim 10^7$}
}
\startdata
\cutinhead{600 days}
 A & 1.2$\times 10^{37}$ & 0.094 & 0.17  & 0.12  &3.4 & 8.3$\times 10^{-3}$\\
 B & 1.1$\times 10^{36}$ & 0.37  & 0.07  & 0.11  &6.6 & 9.4$\times 10^{-5}$\\
 C & 5.6$\times 10^{36}$ & 0.15  & 0.16  & 0.22  &8.1 & 3.7$\times 10^{-3}$\\
 D & 5.9$\times 10^{35}$ & 0.044 & 0.055 & 0.17  &19  & 4.0$\times 10^{-4}$\\
 E & 2.5$\times 10^{37}$ & 0.12  & 0.33  & 0.14  &3.9 & 1.9$\times 10^{-2}$\\
 F & 2.7$\times 10^{36}$ & 0.17  & 0.15  & 0.17  &5.9 & 5.6$\times 10^{-3}$\\
\cutinhead{2500 days}
 A & 3.9$\times 10^{36}$ &  0.018 & 0.035 & 0.059  &2.2 & 3.1$\times 10^{-3}$\\
 B & 1.2$\times 10^{35}$ & 0.010 & 6.7$\times 10^{-3}$ & 0.077  &6.9  &6.3$\times 10^{-5}$\\
 C & 2.1$\times 10^{36}$ & 0.016 & 0.029 & 0.054  &2.0  &2.9$\times 10^{-3}$\\
 D & 4.3$\times 10^{34}$ & 0.013  &0.014 & 0.16 &26 & 3.8$\times 10^{-4}$\\
 E & 9.0$\times 10^{36}$ & 0.019 & 0.071 & 0.055 & 1.9  &5.1$\times 10^{-3}$\\
 F & 8.6$\times 10^{35}$ & 0.015&  0.044 & 0.051 & 0.80  &3.6$\times 10^{-3}$
\enddata
\tablenotetext{a}{The critical densities of the forbidden transitions were taken
from the following sources: [\ion{O}{3}] lines from \citet{De-Robertis:1986}
[\ion{O}{2}]$\;\lambda$7325 from \citet{Tsvetanov:1989}. The value for 
[\ion{C}{1}]$\;\lambda$8727 was estimated from the data provided by \citet{Nussbaumer:1979} 
and  \citet{Pequignot:1976}.}
\end{deluxetable}

\end{document}